\documentclass[lettersize,journal]{IEEEtran}
\usepackage{amsmath,amsfonts}
\usepackage{algorithmic}
\usepackage{algorithm}
\usepackage{array}
\usepackage[caption=false,font=normalsize,labelfont=sf,textfont=sf]{subfig}
\usepackage{textcomp}
\usepackage{stfloats}
\usepackage{url}
\usepackage{verbatim}
\usepackage{graphicx}
\usepackage{booktabs}
\usepackage{cite}
\usepackage[switch]{lineno}

\usepackage{multirow}
\usepackage[justification=centering]{caption} 
\usepackage{hyperref}
\hypersetup{colorlinks=true, linkcolor=blue, anchorcolor=blue, citecolor=blue, urlcolor=blue}
\usepackage{xcolor}

\IEEEoverridecommandlockouts
\makeatletter
\def\ps@IEEEtitlepagestyle{%
  \def\@oddfoot{\mycopyrightnotice}%
  \def\@evenfoot{}%
}
\def\mycopyrightnotice{%
  {\footnotesize 
  \begin{minipage}{\textwidth}
  \centering
  2471-285X~\copyright~2021 IEEE. Personal use is permitted, but republication/redistribution requires IEEE permission. \\ See \url{https://www.ieee.org/publications/rights/index.html} for more information.
  \end{minipage}
  }
  \gdef\mycopyrightnotice{}% just in case
}

\begin{document}

% \linenumbers

\title{Explainable COVID-19 Infections Identification and Delineation Using Calibrated Pseudo Labels}

\author{Ming Li, Yingying Fang, Zeyu Tang, Chibudom Onuorah, Jun Xia, Javier Del Ser,~\IEEEmembership{Senior Member,~IEEE}, \\ Simon Walsh, Guang Yang,~\IEEEmembership{Senior Member,~IEEE}
        % <-this % stops a space
        
\thanks{Ming Li, Yingying Fang, Zeyu Tang, and Chibudom Onuorah are with National Heart and Lung Institute, Imperial College London, London, UK. ~\\ \indent Simon Walsh and Guang Yang are with National Heart and Lung Institute, Imperial College London, London, UK, and Royal Brompton Hospital, London, UK. ~\\ \indent Jun Xia is with Shenzhen Second People's Hospital, Shenzhen, China. ~\\ \indent Javier Del Ser is with TECNALIA, Basque Research \& Technology Alliance (BRTA), 48160 Derio, Spain, and the University of the Basque Country, 48013 Bilbao, Spain.}%
\thanks{Simon Walsh and Guang Yang are co-last senior authors of this work.}%
\thanks{Send correspondence to xiajun@email.szu.edu.cn, g.yang@imperial.ac.uk}%
\thanks{This study was supported in part by Euskampus Foundation (COnfVID19), the Basque Government (IT1294-19, KK-2020/00049), the Project of Shenzhen International Cooperation Foundation (GJHZ20180926165402083), the Clinical Research Project of Shenzhen Health and Family Planning Commission (SZLY2018018), the BHF (TG/18/5/34111, PG/16/78/32402), the ERC IMI (101005122), the H2020 (952172), the MRC (MC/PC/21013), the Royal Society (IEC\textbackslash NSFC\textbackslash211235), the Imperial College Undergraduate Research Opportunities Programme (UROP), the NVIDIA Academic Hardware Grant Program, the SABER project supported by Boehringer Ingelheim Ltd, NIHR Imperial Biomedical Research Centre (RDA01), and the UKRI Future Leaders Fellowship (MR/V023799/1).}%% <-this % stops a space
}

\markboth{manuscript submitted to XXX}
%\markboth{IEEE TRANSACTIONS ON EMERGING TOPICS IN COMPUTATIONAL INTELLIGENCE}
{Shell \MakeLowercase{\textit{et al.}}: A Sample Article Using IEEEtran.cls for IEEE Journals}


\maketitle

\begin{abstract}
The upheaval brought by the arrival of the COVID-19 pandemic has continued to bring fresh challenges over the past two years. 
During this COVID-19 pandemic, there has been a need for rapid identification of infected patients and specific delineation of infection areas in computed tomography (CT) images. 
Although deep supervised learning methods have been established quickly, the scarcity of both image-level and pixel-level labels as well as the lack of explainable transparency still hinder the applicability of AI.
Can we identify infected patients and delineate the infections with extreme minimal supervision?
Semi-supervised learning has demonstrated promising performance under limited labelled data and sufficient unlabelled data.
Inspired by semi-supervised learning, we propose a model-agnostic calibrated pseudo-labelling strategy and apply it under a consistency regularization framework to generate explainable identification and delineation results.
We demonstrate the effectiveness of our model with the combination of limited labelled data and sufficient unlabelled data or weakly-labelled data.
Extensive experiments have shown that our model can efficiently utilize limited labelled data and provide explainable classification and segmentation results for decision-making in clinical routine.
The code is available at \url{https://github.com/ayanglab/XAI_COVID-19}.

\end{abstract}


\section{Introduction}
\IEEEPARstart{i}{t} has been more than two years since we have lived under the shadow of coronavirus disease 2019 (COVID-19) pandemic, which is caused by the severe acute respiratory syndrome coronavirus 2. 
At the time of writing, over 386 million people have been diagnosed with COVID-19, and an estimated 5.7 million death cases were confirmed globally according to the World Health Organization.
The virus binds to epithelial cells in the airway tract, from where it further invades the alveolar epithelial cells causing inflammation and scarring in the lung region\cite{hu2021characteristics, rendeiro2021spatial}. 
\par

The reverse transcriptase-polymerase chain reaction (RT-PCR) test is the current mainstream diagnostic tool as it is rapid and easy to perform.
Samples can be collected from respiratory sources of patients using multiple methods, including nasopharyngeal swabs and oropharyngeal swabs, etc. 
Nonetheless, challenges persist as the molecular detection methods can suffer from a high false-negative rate and low sensitivities in part due to low viral load, improper clinical sampling, and the quality of the test kit\cite{ai2020correlation, khatami2020meta}. 
While computed tomography (CT) can successfully detect COVID-19 in asymptomatic patients and patients with false-negative RT-PCR results \cite{xie2020chest}.
Compared to RT-PCR, the CT-based diagnosis can be made directly after the image acquisition by the assessment of radiologists and also indicate the infected areas and severity of the illness \cite{long2020diagnosis}.
Therefore, researchers are intent on exploring a burgeoning interest in CT to assist COVID-19 patient identification.
Several deep learning methods has been explored to diagnose COVID-19 using CT scans  \cite{wu2021covid, cheng2021automated, zhang2021midcan, ucar2020covidiagnosis}.
Detailed reviews of the existing diagnostic models for COVID-19 can be found in \cite{wynants2020prediction, roberts2021common}. 
\par

Albeit deep learning has demonstrated its success in diagnosing different diseases \cite{li2019recurrent, li2020unified, liu2021deep ,li2020mv}, particular challenges still exist in the diagnosis of COVID-19 \cite{roberts2020machine}.
The CT scans of COVID-19 patients share similar typical features with other pneumonia patients, such as the ground-glass opacity (GGO) in the lung periphery and lower lobes, rounded opacity, enlarged intra-infiltrate vessels, and consolidations in subsegmental areas \cite{adhikari2020epidemiology, harmon2020artificial},
making it difficult to distinguish the COVID-19 cases from other community-acquired pneumonia (CAP) cases and leading to the false-positive cases.
Besides, the performance of most COVID-19 diagnostic models heavily relies on a large and unbiased labelled dataset, which is very expensive and time-consuming considering the cost of manual labelling by experts.
Another challenge lies in the explainability of the predictions as the models are expected to assist doctors in clinical routine. 
Ideally, the infections in the CT scans are expected to be localized and support the predictions of those positive cases.
\par

\begin{figure}
    \centering
    \includegraphics[width=0.8\columnwidth]{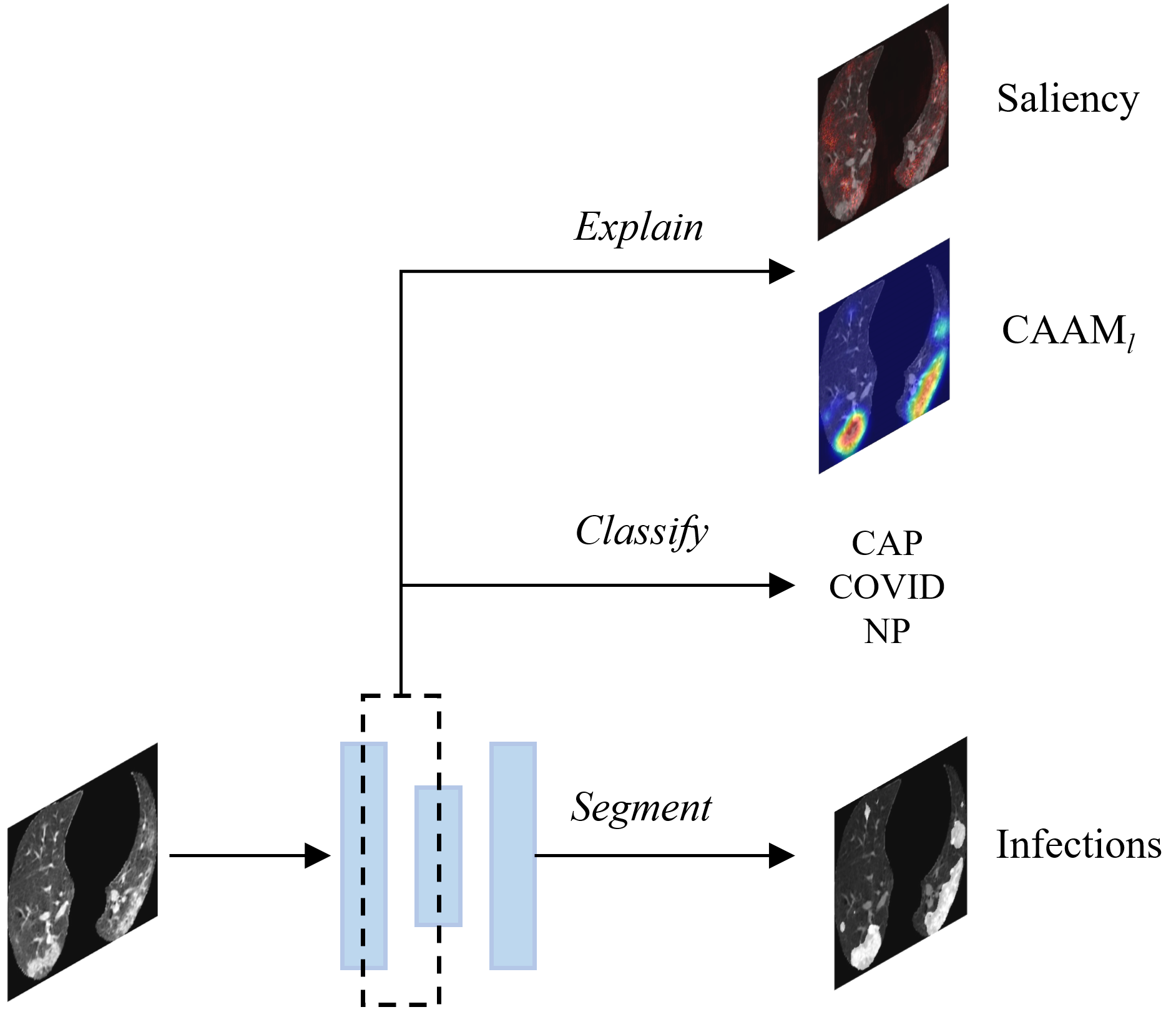}
    \caption{Illustration of our multi-task method which produces classifications, segmentations as well as explainable visualizations at the same time. Here, Integrated Gradients is employed to obtain the saliency map.}
    \label{fig:multi-task}
\end{figure}

To alleviate the aforementioned three challenges: (1) false-positive predictions caused by similar features shared between COVID-19 and CAP cases, (2) the scarcity of both image-level and pixel-level labels, and (3) the lack of explainable transparency,
in this paper, we propose a multi-task explainable identification and delineation method for COVID-19 infections.
As shown in Fig.~\ref{fig:multi-task}, our multi-task method produces classification results (CAP, non-pneumonia (NP), COVID-19) for identification, segmentation results (COVID-19 infections) as well as explainable visualizations (refined class activation map (CAAM$_{l}$), saliency map) as delineation.
We use a multiscale loss function $\mathcal{CAM}_{loss}$ to constrain class-agnostic activation map (CAAM) closer to vanilla class activation map (CAM), enforcing intra-class compactness and inter-class separability, which helps to better distinguish COVID-19 features from CAP features.
By fusing CAAM$_{l}$, saliency map and segmentation together, calibrated pseudo segmentation mask labels (pseudo label) are generated to address the scarcity of pixel-level infection labels.
At the same time, CAAM$_{l}$ and saliency map act as the explainable role, providing auxiliary advice for decision making and verification for segmentation.
The contributions of this paper are summarized as follows:
\begin{itemize}
    \item We propose a multi-task method for explainable identification and delineation of COVID-19 Infections.
    \item The proposed multiscale $\mathcal{CAM}_{loss}$ brings an improvement in classification and segmentation performance.
    \item We design a calibrated pseudo-labelling strategy and apply it under the consistency regularization framework, which only uses limited infection labels.
    \item The proposed method achieves superior performance compared with existing methods, with an accuracy of 77.22\%, and the area under the receiver operating characteristic curve (AUC) of 0.8948 for the classification task as well as the dice score of 85.49\%, and the mean intersection over union (mIoU) score of 76.97\% for segmentation task.
\end{itemize}

\section{Related Work}

\subsection{Supervised COVID-19 Diagnosis}
Currently, most CT-based COVID-19 diagnosis methods are based on supervised learning and obtain promising results.
Wang et al. \cite{wang2020prior} proposed an attention residual learning framework that achieved 93.3\% for accuracy, 87.6\% for sensitivity, and 95.5\% for specificity on a private CT dataset.
% Wang et al. \cite{Wang20212463} developed a supervised joint learning model for COVID-19 classification and lung infection area segmentation, achieving 73.3\% for dice score in segmentation, 94.5\% for accuracy and 94.7\% for sensitivity in classification.
Owais et al. \cite{owais2021multilevel} developed a multilevel deep aggregated boosted network that can diagnose COVID-19. The model was trained and evaluated on six public datasets, achieving 95.38\% for accuracy, 92.53\% for specificity, and  98.14\% for sensitivity. 
% Cheng et al. \cite{Cheng2021} proposed a multitask-learning based autoencoder framework with deep supervision to identify COVID-19 from CT scans, achieving an overall accuracy of 89.54\%.
Zhang et al. \cite{zhang2021midcan} introduced an explainable multi-input deep convolution attention network for COVID-19 classification, achieving a sensitivity of 98.1\%, a specificity of 98.0\%, and an accuracy of 98.0\% on 86 cases.
Wu et al. \cite{wu2021jcs} developed a joint classification and segmentation framework (JCS) for real-time COVID-19 diagnosis using CT scans. They utilized attention mechanism and deep supervision in the classification and segmentation branch, achieving 95.0\% sensitivity and 93.0\% specificity for classification task and 95.9\% dice score for segmentation task.
\par

As mentioned in the introduction, the scarcity of both image-level and pixel-level labels has all these supervised methods over a barrel.
Thus, we design a calibrated pseudo-labelling strategy and apply it under the consistency regularization framework to address the scarcity of labels, see more details in \ref{pseudolabel}.

\subsection{Semi-Supervised Classification and Segmentation}
Existing studies are limited due to the shortage of labelled CT images since manually labelling can be costly and time-consuming.
To combat this, various semi-supervised methods have been investigated to carry out infection detection without relying on large labelled datasets.
Aviles et al.\cite{aviles2022graphxcovid} proposed a graph-based semi-supervised COVID-19 classification framework. An optimization model for graph diffusion was introduced, which reinforced the natural relation between the smaller labelled dataset and the unlabelled dataset. The framework was evaluated on a chest X-ray dataset, achieving 95\% accuracy. 
Fan et al. \cite{fan2020inf} developed a COVID-19 lung infection segmentation model for automatic identification of infected regions in CT scans. A semi-supervised framework based on a randomly selected propagation strategy was used to reduce the reliance on a large labelled dataset. 
Besides, Haque et al. \cite{haque2021generalized} introduced a multi-task semi-supervised learning method (MultiMix) for the classification of pneumonia and segmentation of the lungs in chest X-ray images.
They also preserved explainability in the lung area through a saliency bridge,
but the saliency map they generated mostly focused on the lung periphery and did not match the lung segmentation results very well.
\par

Though semi-supervised methods show promising potential, most of them only use vanilla CAM or inferior methods as explanations, which are not enough for explainable transparency in the prediction of their models.
Hence, we employ multiple explainable methods, i.e., CAAM$_{l}$ and saliency map to provide better explanations for our model, see more details in \ref{explain_two_methods}.

\subsection{Explainable AI with COVID-19}
The mainstream explanation methods used by existing COVID-19 diagnostic models are CAM-based methods \cite{haque2021generalized, hu2020weakly, tabik2020covidgr, aviles2022graphxcovid, wu2021jcs, yang2022unbox}.
CAM was first introduced in 2016 by Zhou et al.\cite{zhou2016learning}, 
a global average pooling layer was added after the last convolutional layer, which helped localize discriminative regions for a particular class. 
A more generalized gradient-weighted CAM was proposed in \cite{selvaraju2017grad}, which can produce a visual explanation for arbitrary model architecture. 
However, the CAM-based methods only focused on the most discriminative areas while neglecting subtle regions and returned coarse-grained explanation maps as the features it used had gone through a sequence of downsamplings. 
The saliency map \cite{zeiler2014visualizing} can produce fine-grained explanation maps with high resolution, but it was very sensitive to the input data and may become unreliable, for instance, the preprocessing procedure (e.g., normalization) can make undesirable changes in the derived saliency map \cite{zhu2017soft, kindermans2019reliability}.
To alleviate these drawbacks, Hu et al.\cite{hu2020weakly} proposed a weakly supervised multiscale joint saliency method for COVID-19 infection detection and classification.
Another novel approach introduced by Baek et al.\cite{baek2020psynet} was to sum up feature maps directly, which generated a class-agnostic activation map (CAAM), indicating the spatial distribution of the embedded features. 
Compared with the vanilla CAM, CAAM showed larger activated areas and richer features but was more error-prone in classification due to the redundant feature representations. 
Thus, Wang et al.\cite{wang2021towards} constructed a loss function CAM-loss that effectively reinforced intra-class compactness and inter-class separability by constraining CAAM closer to CAM.
\par

\section{Method}
\begin{figure}[!t]
    \centering
    \includegraphics[width=1.\columnwidth]{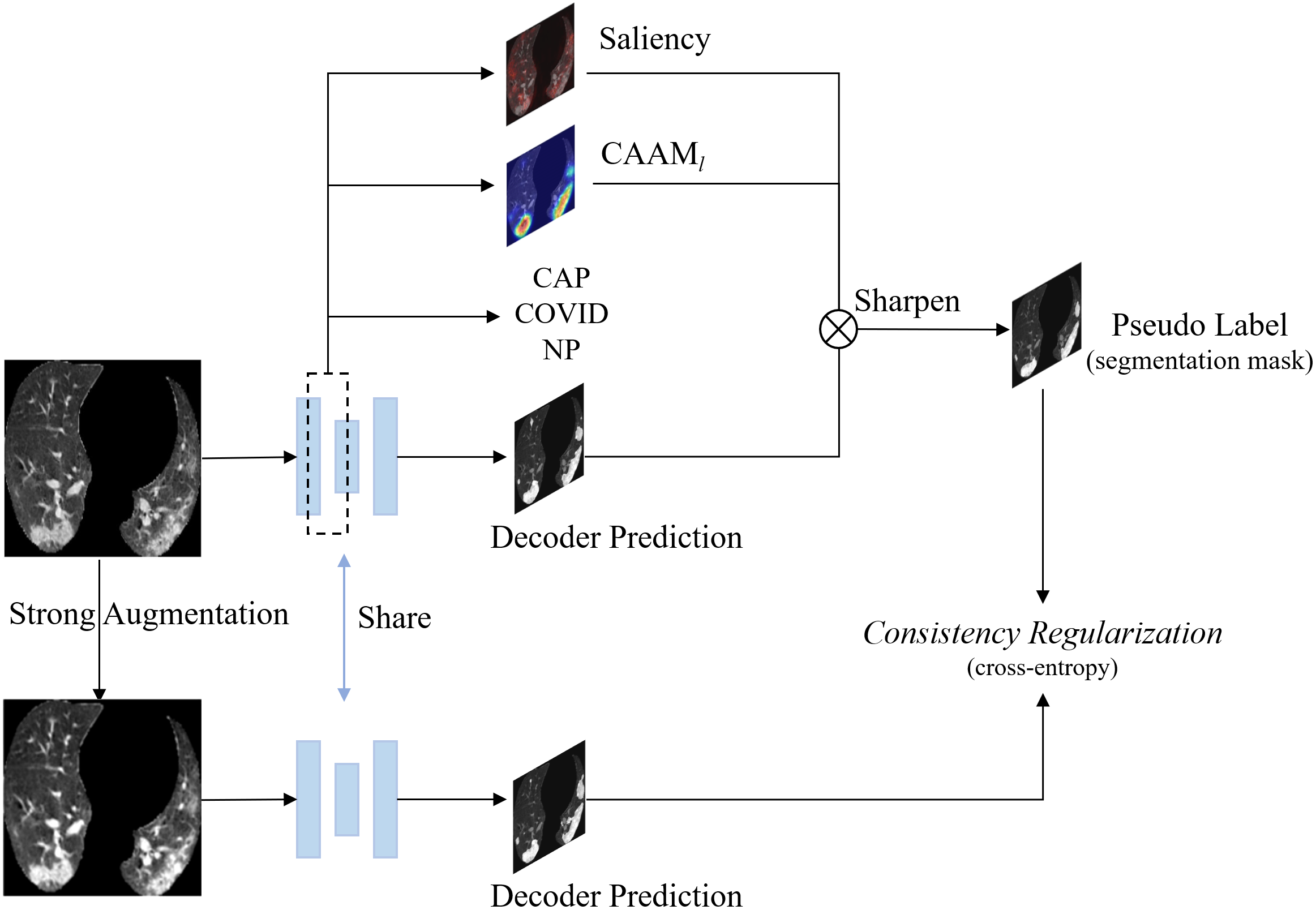}
    \caption{Overview of the framework. Given one unlabelled CT slice, it is fed into the network to obtain a saliency map, CAAM$_{l}$, classification result as well as the decoder prediction.
    Then a calibrated pseudo label is generated via a sharpen combination module.
    Meanwhile, a strongly augmented version of the given unlabelled CT slice goes through the network and gets its decoder prediction.
    Next, we enforce consistency regularization on this decoder prediction to match the calibrated pseudo label via a pixel-wise cross-entropy loss.}
    \label{fig:workflow}
\end{figure}

To assist clinicians in identifying and delineating COVID-19 infections with explainability under extreme minimal supervision, we propose a semi-supervised multi-task method using limited labelled data and sufficient unlabelled data.
The overview of the framework is depicted in Fig.~\ref{fig:workflow}.
% Given one unlabelled CT slice, it is fed into the network to obtain saliency map, CAAM$_{l}$, classification results as well as the decoder prediction.
% Then a calibrated pseudo label is generated via a sharpen combination module.
% Meanwhile, a strongly augmented version of the given unlabelled CT slice goes through the network and gets its decoder prediction.
% Next, we enforce consistency regularization on this decoder prediction to match the calibrated pseudo label via a pixel-wise cross-entropy loss.
Following, we describe more details of the proposed method.

\subsection{Prerequisite: Lung Area Extraction}
By suppressing the disturbance of the non-lung area \cite{hu2020weakly, ye2021explainable}, we can facilitate the COVID-19 infection classification and segmentation with higher accuracy as well as better explainability.
We follow the procedure described in \cite{hu2020weakly}, using the TCIA open dataset \cite{yang2018autosegmentation} to train a lung area segmentation model.
The trained lung area segmentation model is then utilized to segment the entire lung structure of the COVID-19, CAP, and NP patients included in our private dataset.

\subsection{Backbone Architecture}
As shown in Fig.~\ref{fig:camloss}, we use VGG architecture \cite{simonyan2014deep} as our backbone.
Five convolutional blocks (i.e., Conv1, Conv2, Conv3, Conv4, and Conv5) extract multiscale features, from bottom to top, each block consists of 2, 2, 3, 3, 3 convolutional layers, respectively.
Every convolutional layer is a combination of convolution layers with $3\times3$ kernels, instance normalization, and Leaky Rectified Linear Units.
There are 32, 64, 128, 256, and 256 filters in the corresponding convolutional blocks and 3 filters in the $1\times1$ convolution layer.
Each convolutional block is followed by a max-pooling layer with a kernel size of $2\times2$.

\subsection{Multiscale \texorpdfstring{$\mathcal{CAM}_{loss}$}{\space} for Classification}
\begin{figure*}[!ht]
    \centering
    \includegraphics[width=1.6\columnwidth]{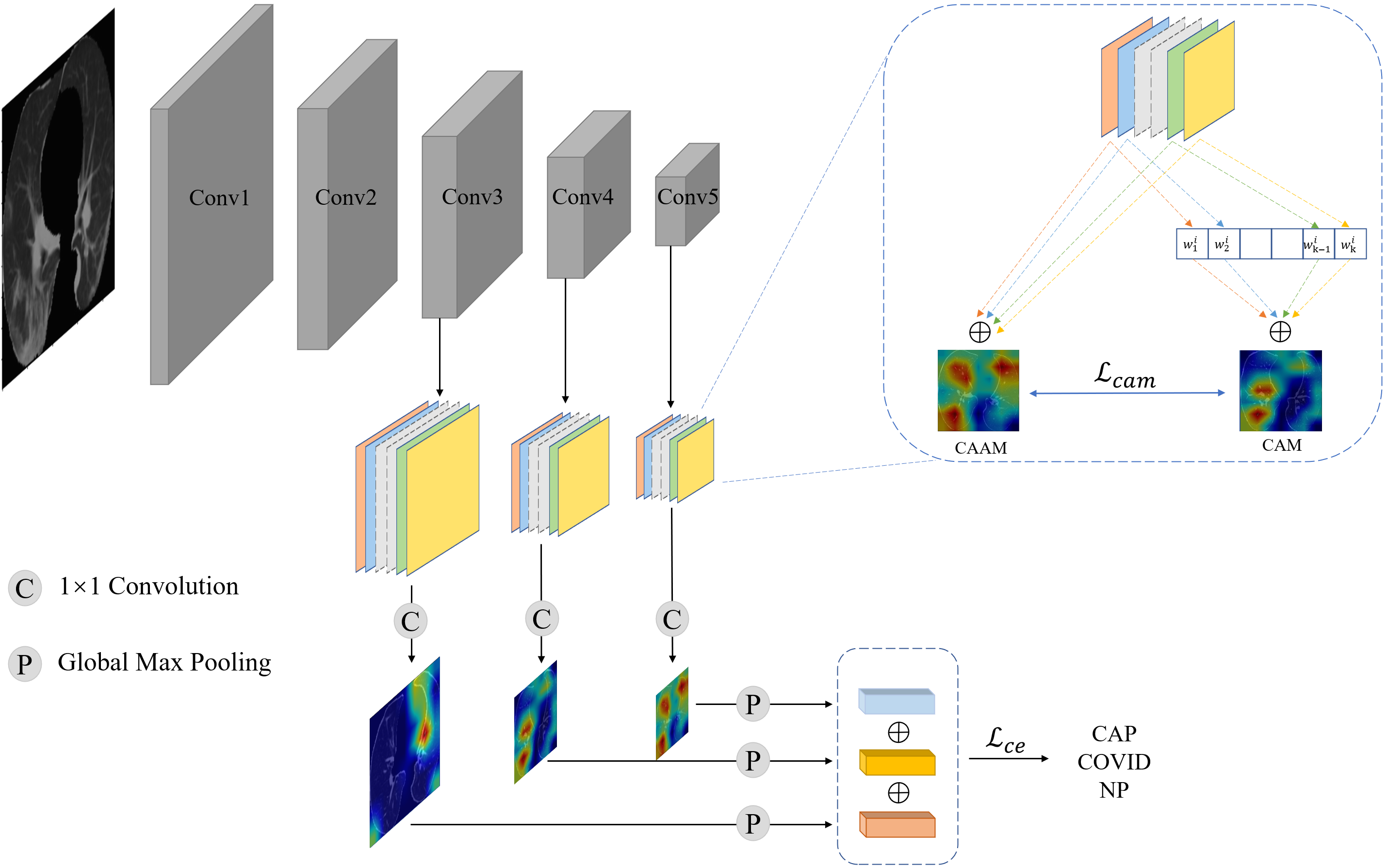}
    \caption{Backbone architecture and $\mathcal{CAM}_{loss}$. The dashed box area (encoder and classifier) in Fig.~\ref{fig:multi-task} and Fig.~\ref{fig:workflow}.}
    \label{fig:camloss}
\end{figure*}

Considering the COVID-19 infections appear to be small in mild cases, while in severe cases they scatter over a large area \cite{pan2020time, shi2020radiological}.
We follow \cite{hu2020weakly} to implement a multiscale learning scheme to cope with the variation of infection location and size by learning different features in various scales.
The mid-level layer can learn small lesions (e.g., GGO), but is not able to capture larger scattered lesions due to limited receptive field.
While the higher-level layers with larger receptive fields are capable to detect larger scattered and diffuse patchy-like lesions.
As depicted in Fig.~\ref{fig:camloss}, multiscale feature maps from Conv3, Conv4, and Conv5 are mapped down to the class score maps by $1\times1$ convolution.
Followed with a Global Max Pooling operation, the class score maps are transformed into categorical score vectors.
By summing up three categorical score vectors, we can make a final prediction with a $\textrm{Softmax}$ activation function.
Then a cross-entropy loss 
\begin{equation}
    \mathcal{L}_{\mathrm{ce}}=-\sum^{C}_{i} y_{i} \mathrm{log}(x_{i})
\end{equation} 
is utilized for the classification task, where $C$ is the number of classes, $y_{i}$ and $x_{i}$ are the ground truth (class) and the predicted class score of class $i$, respectively.
% \begin{equation}
%     \mathcal{L}_{\mathrm{ce}}=-\log \frac{e^{z_{i}}}{\sum^{C}_{j} e^{z_{j}}}
% \end{equation} 
% is utilized for the classification task, where $C$ is the class number, $z_{i}$ is the class score of class $i$.
\par

However, as aforementioned, the COVID-19 infections share very similar common features with CAP, for example, the GGO as well as the airspace consolidation, both of them distribute bilaterally in the lung periphery.
Thus \cite{hu2020weakly} only achieved suboptimal performance on telling COVID apart from CAP.
To better distinguish COVID-19 from CAP, we introduce a multiscale $\mathcal{CAM}_{loss}$ inspired by \cite{wang2021towards, dong2018holistic}.

For a given image, we take class score maps from Conv5 as an example.
The vanilla CAM for class $i$ is a weighted sum of the feature maps, which can be formulated as 
\begin{equation}
\textrm{CAM}_{i}=\sum_{k} w_{k}^{i} f_{k},
\end{equation}
where $f_{k}$ is the $k^{th}$ feature map, and $w^{i}_{k}$ represents the weight corresponding to class i for $f_{k}$.
While CAAM is a direct summation of all feature maps, it is given by 
\begin{equation}
\textrm{CAAM}=\sum_{k} f_{k}.
\end{equation}

For a target category, CAAM generally shows richer features and larger areas than CAM$_{i}$ which only localizes partial regions of interest.
However the redundant features in CAAM may result in higher confidence scores of non-target categories and lead to misclassification.
To enforce intra-class compactness and inter-class separability, we constrain CAAM closer to CAM$_{i}$ by a loss $\mathcal{L}_\mathrm{cam}$. 
The $l_{1}$ distance is employed in $\mathcal{L}_\mathrm{cam}$ to measure the distance between CAAM and CAM$_{i}$, driving the backbone to capture target category features and suppress features of the non-target categories.
This loss is defined as
\begin{equation}
\mathcal{L}_{\mathrm{cam}}=\left\|\textrm{CAAM}^{\prime}-\textrm{CAM}^{\prime}_{i}\right\|_{l_{1}},
\end{equation}
where CAAM$^{\prime}$ and CAM$^{\prime}_{i}$ are min-max normalized CAAM and CAM$_{i}$.
Besides Conv5 level, we also apply $\mathcal{L}_\mathrm{cam}$ on Conv3 and Conv4 level to formulate a multiscale $\mathcal{CAM}_{loss}$.
For the classification task, the final formal multiscale $\mathcal{CAM}_{loss}$ can be formulated as
\begin{equation}
\mathcal{CAM}_{loss}=\sum_{s} \alpha_{s}\mathcal{L}^{s}_{\mathrm{cam}}+\mathcal{L}_{\mathrm{ce}},
\label{cam_loss}
\end{equation}
where $\mathcal{L}^{s}_\mathrm{cam}$ indicates $\mathcal{L}_\mathrm{cam}$ at $s^{th}$ scales ($s=1,\ldots,3$, i.e., the Conv3, Conv4, Conv5 levels), and $\alpha_{s}$ represents the corresponding combination ratio.

\subsection{Saliency Map}
How to generate a satisfactory pseudo label for segmentation? 
The most straightforward way is simply employing the confidence thresholded decoder prediction from a trained segmentation model \cite{sohn2020fixmatch}.
However, this kind of pseudo label is barely desirable when training with limited labelled data and thus yield inferior performance \cite{zou2020pseudoseg}.
Intuitively, we can solve this issue by combining multiple sources of prediction maps with the decoder prediction as compensation and further generating a well-calibrated pseudo label.
Here, we employ the refined class activation map CAAM$_{l}$ and saliency map as the compensation.
\par

The CAAM$_{l}$ is able to produce a more discriminative and locally consistent mask for segmentation but is still barely satisfactory in delineating crisp boundaries of the infection areas since it is obtained from low-resolution feature maps.
While it is possible to generate high-resolution maps with subtle boundaries by saliency maps \cite{simonyan2014deep}.
Here we employ Integrated Gradients \cite{sundararajan2017axiomatic} to obtain the saliency map. 
Since it is applicable to a variety of deep networks and can be implemented using a few calls to the gradients operator and has a strong theoretical justification.
\par

Formally, let function $F$ represent the network and suppose we have a flattened input image denoted as $x=(x_{1},\ldots,x_{n}) \in \mathbb{R}^{n}$ as well as a baseline input (black image filled with zeros in our case) $x^{\prime}=(x^{\prime}_{1},\ldots,x^{\prime}_{n}) \in \mathbb{R}^{n}$ where n is the number of pixels.
Then $a_{i}$, the contribution of individual pixel $x_{i}$ to the prediction $F(x)$ relative to the baseline input $x^{\prime}_{i}$, can be written as a vector $\mathbf{A}_{F}(x, x^{\prime}) = \frac{\partial F(x)}{\partial x} = (a_{1},\ldots, a_{n}) \in \mathbb{R}^{n}$, which is also the gradient of $F(x)$ at $x$.
% where $a_{i}$ is the contribution of $x_{i}$ to the prediction $F(x)$.
Next, integrated gradients can be obtained by cumulating all gradients computed at each point along the straight-line path (in $\mathbb{R}^{n}$) from the baseline $x^{\prime}$ to the input $x$.
For the $i^{th}$ dimension of an input $x$ and baseline $x^{\prime}$, the integrated gradient is given by
\begin{equation}
\textrm{InGrads}_{i}(x)=\left(x_{i}-x_{i}^{\prime}\right) \times \int_{\alpha=0}^{1} \frac{\partial F\left(x^{\prime}+\alpha\left(x-x^{\prime}\right)\right)}{\partial x_{i}} d \alpha,
\end{equation}
where $\frac{\partial F(x)}{\partial x_{i}}$ is the gradient of $F(x)$ along the $i^{th}$ dimension.
For the sake of computational efficiency, the integral of integrated gradients can be approximated via a summation as
\begin{equation}
\textrm{InGrads}_{i}(x) \approx \left(x_{i}-x_{i}^{\prime}\right) \times \sum_{g=1}^{t} \frac{\partial F\left(x^{\prime}+\frac{g}{t} \times\left(x-x^{\prime}\right)\right)}{\partial x_{i}} \times \frac{1}{t},
\end{equation}
where $t$ is the number of steps in Riemann's approximation of the integral and in practice, we compute the gradient in a for loop over the set of inputs (i.e., $g=1,\ldots,t$).
To obtain the saliency map, we directly use integrated gradients computed from the input CT scan and the corresponding multiscale classification output.
\par

\begin{figure*}[!th]
    \centering
    \includegraphics[width=1.9\columnwidth]{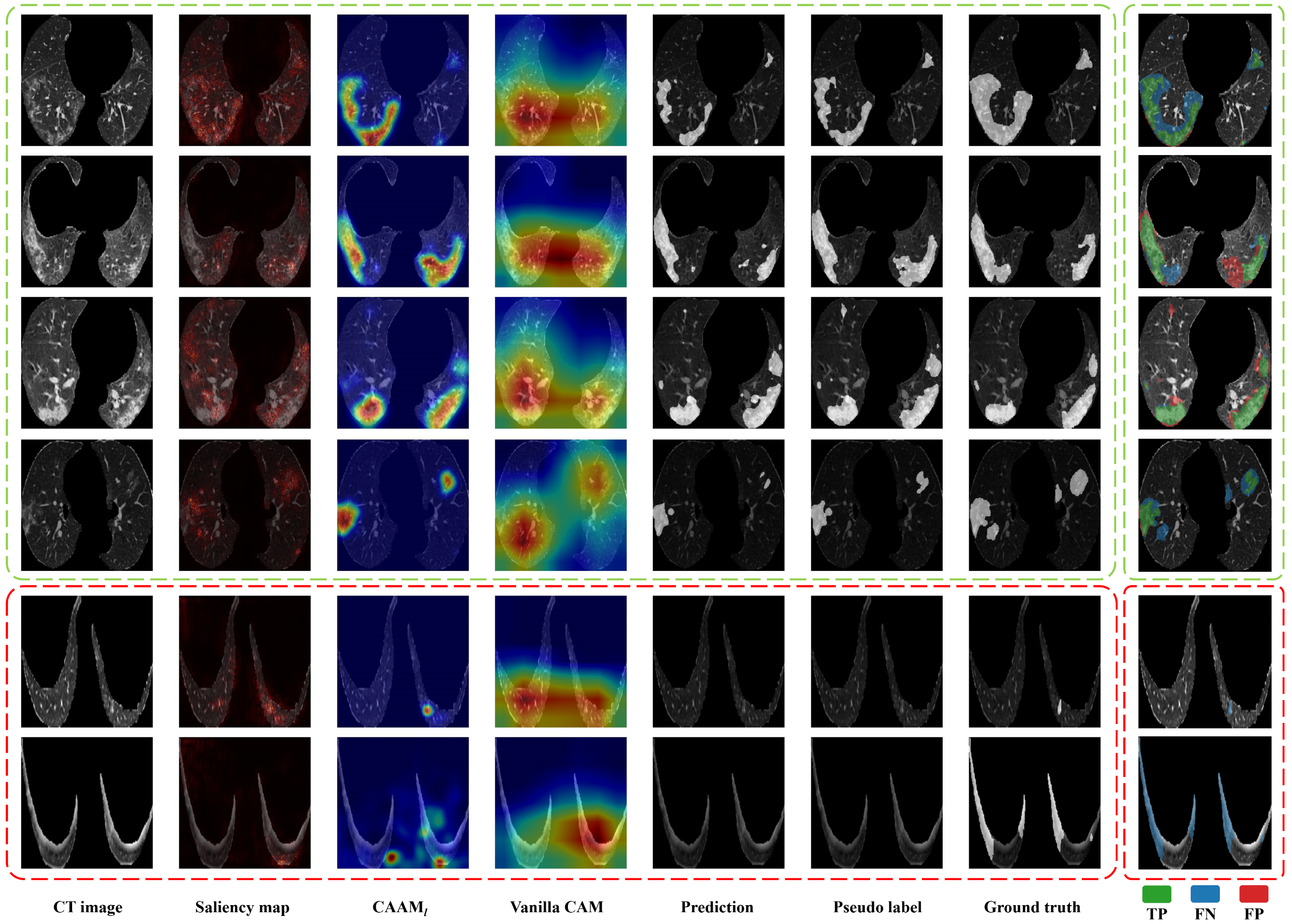}
    \caption{Visualizations of the CT images and the corresponding saliency maps, CAAM$_{l}$, vanilla CAM, decoder prediction, pseudo label as well as the ground truth (infection mask). By combining saliency map, CAAM$_{l}$, and decoder prediction, we get the calibrated pseudo label. The last column shows the true positive (TP), false negative (FN) and false positive (FP) visualizations in different colors. (Top four rows in the green dashed box: good cases; Bottom two rows in the red dashed box: bad cases)}
    \label{fig:maps_all}
\end{figure*}

\subsection{Pseudo Label Calibration for Segmentation}\label{pseudolabel}
Given the CAAM$_{l}$ $c$, the saliency map $s$, and the decoder prediction $p$, how to fuse them together for a better pseudo label $\tilde{y}$? 
Inspired by the fusion method proposed in PseudoSeg \cite{zou2020pseudoseg}, we design a sharpen combination module $S$ to calibrate the pseudo label.
The $S$, which consists of two key operations: $\textrm{Norm}$ and $\textrm{Sharpen}$, can be formulated as
\begin{equation}
\begin{split}
S(c,s,p)=\textrm{Sharpen}&\left.\Bigg\{\lambda \textrm{Softmax}\left(\frac{c}{\textrm{Norm}(c,s,p)}\right) \right.\\ %\left.   % \right. \\ \left. for wrapping line in equation
&\left.+\mu \textrm{Softmax}\left(\frac{s}{\textrm{Norm}(c,s,p)}\right) \right.\\ %\left.
&\left.+\nu \textrm{Softmax}\left(\frac{p}{\textrm{Norm}(c,s,p)}\right), T\right.\Bigg\}
\end{split},
\end{equation}
where $\lambda + \mu + \nu=1$.
In view of $c$, $s$, and $p$ could have quite different degrees of overconfidence since they come from different decision mechanisms.
We use a normalization operation 
\begin{equation}
\textrm{Norm}(c,s,p)=\sqrt{\sum_{i}^{|c|}\left(c_{i}^{2}+s_{i}^{2}+p_{i}^{2}\right)},
\end{equation}
to ease the overconfident probability with $\textrm{Softmax}$ and thus avoid the unbalanced term in $S$.
Next, the distribution sharpening operation 
\begin{equation}
\textrm{Sharpen}(a,T)_{i}= \frac{a_{i}^{1 / T}}{\sum_{j}^{C} a_{j}^{1 / T}},
\end{equation}
adjusts the categorical distribution \cite{berthelot2019mixmatch} and produces the final calibrated pseudo label $\tilde{y}$.
Here $T$ is a hyperparameter which is called the temperature scalar, it is used to control the categorical distribution.
\par

For segmentation, our training procedure contains a supervised loss $\mathcal{L}_\mathrm{s}$ for pixel-level labelled data $\mathcal{D}_\mathrm{s}$ and a consistency regularization loss $\mathcal{L}_\mathrm{u}$ for unlabelled data $\mathcal{D}_\mathrm{u}$.
The $\mathcal{L}_\mathrm{s}$, a weighted binary cross-entropy loss function, is given by
\begin{equation}
\mathcal{L}_{\mathrm{s}}=- y \times \mathrm{log}(F_{\theta}(x)) - w(1-y)\times \mathrm{log}(1-F_{\theta}(x)),
\label{loss_supervised}
\end{equation}
where $\theta$ represents the learnable parameters of the network function $F$, $x$ is the input, $y$ is the corresponding ground truth (infection mask), and $w$
% $w=\frac{\textrm{number of positive pixels}}{\textrm{number of negative pixels}}$ 
is the weight coefficient set to cope with the imbalanced problem.
With regards to the unsupervised counterpart, our method generates a pseudo label $\tilde{y}$ for every strongly augmented unlabelled data $x$ from $\mathcal{D}_{u}$.
Following the consistency training technique \cite{xie2020unsupervised}, we design a consistency regularization loss based on weighted binary cross-entropy loss function, which can be written as
\begin{equation}
\mathcal{L}_{\mathrm{u}}=- \tilde{y} \times \mathrm{log}(F_{\theta}(\Theta(x))) - w(1-\tilde{y})\times \mathrm{log}(1-F_{\theta}(\Theta(x))),
\label{loss_unsupervised}
\end{equation}
where $\Theta$ indicates the strong data augmentation operation.
\par

Finally, the training objective consists of the classification loss $\mathcal{CAM}_{loss}$, the segmentation loss $\mathcal{L}_{\mathrm{s}}$, and the consistency regularization loss $\mathcal{L}_{\mathrm{u}}$.
All of the loss terms are optimized jointly with corresponding coefficients (i.e., $\beta\mathcal{CAM}_{loss}+\gamma\mathcal{L}_{\mathrm{s}}+\eta\mathcal{L}_{\mathrm{u}}$), while $\mathcal{L}_{\mathrm{u}}$ is incorporated only after the $\mathcal{CAM}_{loss}$ and $\mathcal{L}_{\mathrm{s}}$ are stable (i.e., until the classification and segmentation results become acceptable, which means the results are comparable with baseline models).
The training process is summarized in Algorithm~\ref{algorithm_1}.

\begin{table*}[!ht]
\begin{center}
\caption{Detail descriptions and imaging parameters of the CT systems of both our private dataset and open dataset.}
\label{tab_data}
\begin{tabular}{| c | c | c | c | c | c |}
\hline
\multirow{2}{*}{Centre} & \multicolumn{3}{c|}{\multirow{2}{*}{SZSH}} & \multirow{2}{*}{WHRCH} & \multirow{2}{*}{Open Dataset\cite{majun2020opendata}} \\
 & \multicolumn{3}{c|}{} & & \\
\hline
Category & NP & CAP & COVID-19 & COVID-19 & COVID-19 \\
\hline
Infection label (pixel-level) & $\times$ & $\times$ & $\times$ & $\times$ & \checkmark \\
\hline
Cases  & $115$ & $124$ & $12$ & $108$ & $20$ \\
\hline
Slice number (resampled) & $23,931$ & $22,989$ & $1,939$ & $11,834$ & $4,151$ \\
\hline
\multirow{2}{*}{Scanner} & \multicolumn{2}{c|}{SIEMENS} & GE  & SIEMENS & \multirow{2}{*}{-} \\
 & \multicolumn{2}{c|}{SOMATOM Emotion} & revolution 256 & SOMATOM go.Now16 &  \\
\hline
Tube voltage & \multicolumn{2}{c|}{$110~KV$} & $120~KV$ & $130~KV$ & - \\
\hline
Slice thickness & \multicolumn{2}{c|}{$1.2~mm$} & $0.625~mm$ & $0.7~mm$ & - \\ 
\hline
Reconstructed slice thickness & \multicolumn{2}{c|}{$1.5~mm$} & $2~mm$ & $1~mm$ & $(1, 1.5, 2, 4, 6)~mm$ \\ % 1~mm, 1.5~mm, 2~mm, 4~mm, 6~mm
\hline
Pitch & \multicolumn{2}{c|}{$1.2$} & $1.375$ & $1.5$ & - \\
\hline 
Matrix & \multicolumn{2}{c|}{$512\times512$} & $512\times512$ & $512\times512$ & $512\times512$ \\
\hline 
Field of view & \multicolumn{2}{c|}{$260~mm\times260~mm$} & $400~mm\times400~mm$ & $350~mm\times350~mm$ & - \\
\hline 
Automatic tube current modulation & \multicolumn{2}{c|}{$70~mAs$} & $150~mAs$ & $50~mAs$ & - \\
\hline 
\end{tabular}
\end{center}
% \vspace{-0.3cm}
\end{table*}

\begin{algorithm}[h]
        \caption{Training process} 
        \hspace*{0.02in} {\bf Initialization:} 
        the parameters of model $\theta$; \\
        \hspace*{0.02in} {\bf Optimization:} 
        \begin{algorithmic}[1]
            \FOR{number of training iterations} 
            \STATE Calculate CAM, CAAM, and Saliency Map
            \IF{performance \textless ~baseline}
            \STATE Update $\theta$ with $\nabla _{\theta}(\mathcal{CAM}_{loss}~^{\ref{cam_loss}},~\mathcal{L}_{s}~^{\ref{loss_supervised}})$
            \ENDIF
            \STATE Update $\theta$ with $\nabla _{\theta}(\mathcal{CAM}_{loss}~^{\ref{cam_loss}},~\mathcal{L}_{s}~^{\ref{loss_supervised}},~\mathcal{L}_{u}~^{\ref{loss_unsupervised}})$
            \ENDFOR
            \RETURN optimal $\theta$
        \end{algorithmic}
        \label{algorithm_1}
\end{algorithm}

\begin{table*}
\begin{center}
\caption{Details of the data split (numbers are shown as X cases (Y slices)). In the labelled training set, we have patient-level category labels and pixel-level infection labels (NP/CAP slices have all zero segmentation masks and are uniformly sampled from all NP/CAP cases in the full training set), while in the unlabelled training set, there are no labels available at all during training.}
\label{data_split}
\begin{tabular}{lcccccc}
\toprule
\multirow{2}{*}{Sets} & \multirow{2}{*}{Total} & \multicolumn{3}{c}{Private} & & \multicolumn{1}{c}{Open} \\
\cline{3-5} \cline{7-7}
 & & CAP & NP & COVID-19 & & COVID-19 \\
\midrule
Full dataset & 
379 (64,844) & 124 (22,989) & 115 (23,931) & 120 (13,773) & & 20 (4,151) \\
\midrule
Full training set &
192 (32,779) & 64 (12,250) & 55 (11,025) & 60 (6,689) & & 13 (2,815) \\
Full testing set &
187 (32,065) & 60 (10,739) & 60 (12,906) & 60 (7,084) & & 7 (1,336) \\
\midrule
Labelled training set & 
132 (7,815) & 64 (2,500) & 55 (2,500) & - & & 13 (2,815) \\
Unlabelled training set & 
179 (24,964) & 64 (9,750) & 55 (8,525) & 60 (6,689) & & - \\
\midrule
Classification testing set &
187 (32,065) & 60 (10,739) & 60 (12,906) & 60 (7,084) & & 7 (1,336) \\
Segmentation testing set &
7 (1,336) & - & - & - & & 7 (1,336) \\
\bottomrule
\end{tabular}
\end{center}
% \vspace{-0.3cm}
\end{table*}

\subsection{Explanation}\label{explain_two_methods}
Our method employs the CAAM$_{l}$ and the saliency map as explanations to increase the explainable transparency of the identification and delineation of the COVID-19 infections.
As aforementioned, the CAAM$_{l}$, refined by $\mathcal{CAM}_{loss}$, is capable to perform better in suppressing non-target features and capturing target features than vanilla CAM.
It learns more globally discriminative and locally consistent features of the lesion area.
While the multiscale saliency map can provide more subtle boundaries in the lesion area.
Both the CAAM$_{l}$ and the saliency map were refined together with the decoder prediction by the sharpen combination module during the learning procedure.
Finally, the trained model is able to provide convincing explanation results.
Fig.~\ref{fig:maps_all} shows some visualization results (both good and bad cases).

\section{Experiments and Discussion}
In this section, we detail the conducted experiments to validate our proposed method.
We start by specifying the data, implementation, and evaluation details.
Next, we conduct comparison studies in the supervised setting and the semi-supervised setting, respectively.
Lastly, we evaluate our model design choices through various ablation studies.

\subsection{Data Description}\label{data_datail}
We use an open dataset \cite{majun2020opendata} with pixel-level annotations of lungs and infections made by experts to train our model.
This open dataset contains 20 CT scans (4,151~slices) of patients diagnosed with COVID-19 as well as annotations made by experts.
Besides, we collected 359 patient CT scans (60,693~slices) with patient-level labels from the Hospital of Wuhan Red Cross Society (WHRCH) and Shenzhen Second Hospital (SZSH) between September 2016 and March 2020.
Our private dataset contains 120, 124, 115 chest CT scans of COVID-19, CAP, and NP patients, respectively.
All COVID-19 patients, scanned within 3 days of hospitalisation from December 2019 to March 2020, were tested positive by RT-PCR and confirmed at severe or critical stage according to the diagnosis and treatment guidelines \cite{wang2020diagnosis} of COVID-19 issued by the Chinese National Health Commission.
Both CAP and NP patients were from SZSH, for COVID-19 patients, 108 of them were from WHRCH, the other 12 patients were also from SZSH.
Details are shown in Table \ref{tab_data}.
\par

Specifically, 179 cases (29,964 slices, NP: 55 cases, CAP: 64 cases, COVID-19: 60 cases) from our private dataset and 13 cases (2,815 slices, COVID-19: 13 cases) from the open dataset are combined as the full training set (32,779 slices, NP: 55 cases, CAP: 64 cases, COVID-19: 73 cases).
The remaining 180 cases (30,729 slices, NP: 60 cases, CAP: 60 cases, COVID-19: 60 cases) in our private dataset and 7 cases (1,336 slices, COVID-19: 7 cases) in the open dataset consists of the full testing set (32,065 slices, NP: 60 cases, CAP: 60 cases, COVID-19: 67 cases).
See more details about data split in Tabel~\ref{data_split}.
% \vspace{-0.1cm}

\begin{table*}[!t]
\begin{center}
\caption{Classification performance comparison among different methods. $\dagger$ :supervised, $\star$ :semi-supervised}
\label{classify_performance}
\setlength{\tabcolsep}{3mm}
{
\begin{tabular}{lcccc}
\toprule
Method & Accuracy(\%) & Sensitivity(\%) & Specificity(\%) & AUC \\
\midrule
VGG-16 (baseline)$\dagger$ & 70.65 & 72.93 & 84.65 & 0.8600 \\
JCS \cite{wu2021jcs}$\dagger$ & 71.90 & 74.65 & 85.58 & 0.8548 \\
ResNet-18$\dagger$  & 71.42 & 73.68 & 84.91 & 0.8689 \\
Hu et al.\cite{hu2020weakly}$\dagger$ & 71.05 & 73.45 & 84.93 & 0.8703 \\
Pseudo-Label \cite{lee2013pseudo}$\star$ & 72.31 & 74.76 & 86.02 & 0.8810 \\
MultiMix \cite{haque2021generalized}$\star$ & 72.91 & 74.37 & 85.69 & 0.8701 \\
Ours$\dagger$ & 71.40 & 73.82 & 85.54 & \textbf{0.8851} \\
Ours$\star$ & \textbf{75.13} & \textbf{76.35} & \textbf{86.89} & 0.8748 \\
% Ours (weak) & \textbf{77.22} & \textbf{78.01} & \textbf{87.97} & \textbf{89.48} \\
\bottomrule
\end{tabular}
}
\end{center}
% \vspace{-0.3cm}
\end{table*}

\begin{figure*}[!th]
    \centering
    \includegraphics[width=1.9\columnwidth]{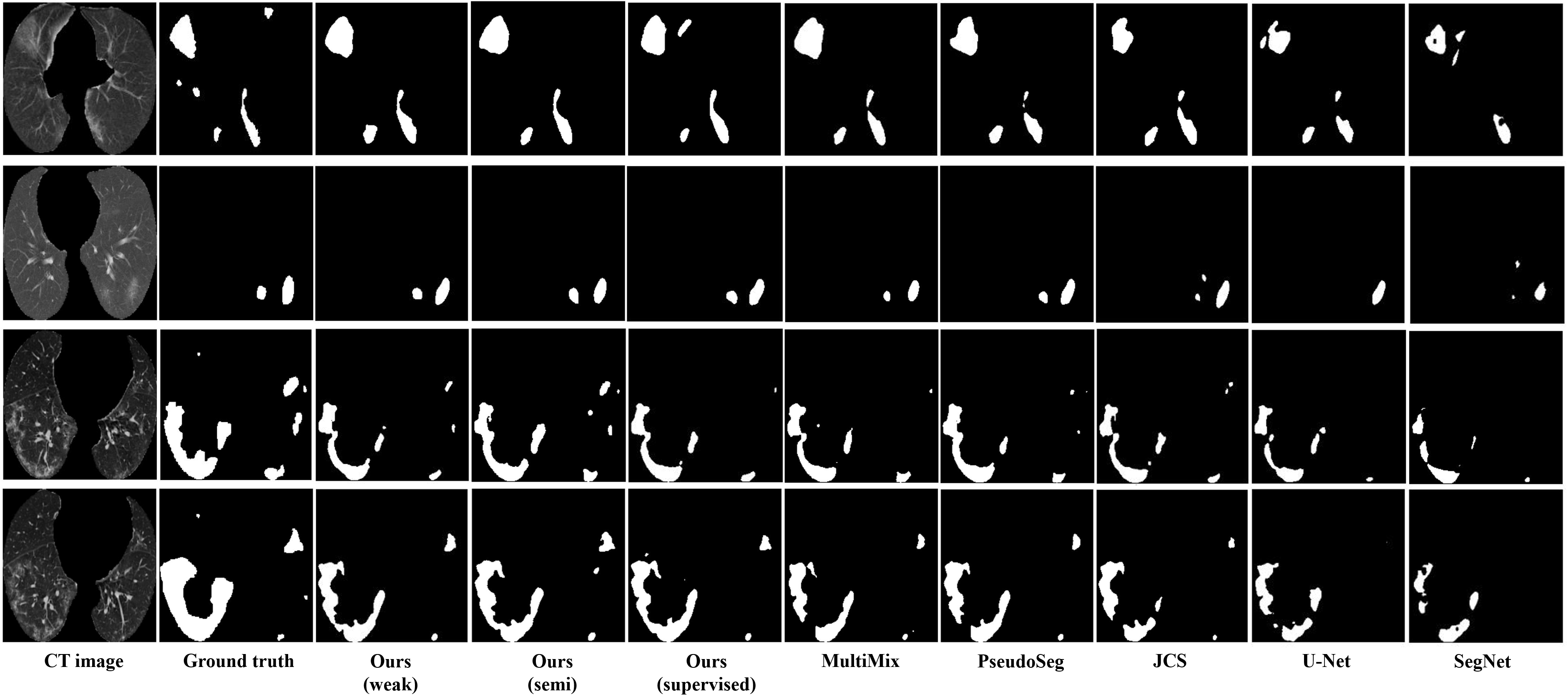}
    \caption{Visual comparisons of different methods for COVID-19 infection lesions segmentation.}
    \label{fig:segs}
\end{figure*}

\begin{table}[!b]
\begin{center}
\caption{Segmentation performance comparison among different methods. $\dagger$ :supervised, $\star$ :semi-supervised}
\label{segmentation_performance}
\setlength{\tabcolsep}{5mm}
{
\begin{tabular}{lcc}
\toprule
Method & Dice(\%) & mIoU(\%) \\
\midrule
SegNet (baseline)$\dagger$ & 75.64 & 67.31 \\
U-Net$\dagger$ & 78.90 & 70.52 \\
JCS \cite{wu2021jcs}$\dagger$ & 79.71 & 71.36 \\ 
PseudoSeg \cite{zou2020pseudoseg}$\star$ & 82.88 & 73.99 \\
MultiMix \cite{haque2021generalized}$\star$ & 84.80 & 76.17 \\
Ours$\dagger$ & 84.53 & 75.90 \\
Ours$\star$ & \textbf{85.49} & \textbf{76.97} \\
% Ours (weak) & 83.67 & 74.92 \\
\bottomrule
\end{tabular}
}
\end{center}
\end{table}

\subsection{Implementation Details and Evaluation Metrics}
All CT slices are resampled to $1\times1\times1~mm$ and resized to $224\times224$ for computational efficiency.
The proposed model is trained for 300 epochs on a Linux workstation with four Nvidia RTX 3090 GPUs using PyTorch 1.10.
We adopt the Adam optimiser ($\beta_{1} = 0.5$, $\beta_{2} = 0.9$) with a initial learning rate $10^{-4}$.
The learning rate is updated by multiplying by 0.1 in every 20 epochs.
We employ contrast adjustment and sharpness adjustment as strong data augmentation methods.
All $\alpha_{s}$ in $\mathcal{CAM}_{loss}$ are set to 5.
The $\lambda$, $\mu$, and $\nu$ in sharpen combination module are 0.3, 0.4, 0.4, respectively.
And $T$ in the sharpening operation is set to 0.5.
As for the coefficients in the final loss function, $\beta$, $\gamma$, and $\eta$ are set to 1, 5, 5 separately.
\par

With regard to evaluation metrics, for classification performance evaluation, we use accuracy, the area under the receiver operating characteristic curve (AUC), sensitivity, and specificity as suggested by \cite{liu2020rethinking}.
For segmentation performance evaluation, we employ two standard metrics, dice score and mean intersection over union (mIoU).

\begin{table*}[!t]
\begin{center}
\caption{Ablation study results (patient-level labels: - indicates semi-supervised setting; \checkmark represents weakly-supervised setting).}
\label{ablation_results}
\begin{tabular}{lcccccccc}
\toprule
\multirow{2}{*}{Method} & \multirow{2}{*}{patient-level labels} & \multicolumn{4}{c}{Classification} & & \multicolumn{2}{c}{Segmentation} \\
\cline{3-6} \cline{8-9}
 & & Accuracy(\%) & Sensitivity(\%) & Specificity(\%) & AUC & & Dice(\%) & mIoU(\%) \\
\midrule
w/o cam loss & - & 73.66 & 75.53 & 86.26 & 0.8633 &  & 80.53 & 71.62 \\
w/o saliency map & - & 73.80 & 74.21 & 85.81 & 0.8634 &  & 84.09 & 75.41 \\
w/o sharpen & - & 74.65 & 75.90 & 86.63 & 0.8597 &  & 82.14 & 73.28 \\
w/o multiscale & - & 74.78 & 76.23 & 86.72 & 0.8661 &  & 82.69 & 73.90 \\
full & - & \textbf{75.13} & \textbf{76.35} & \textbf{86.89} & \textbf{0.8748} &  & \textbf{85.49} & \textbf{76.97} \\
\midrule
w/o cam loss & \checkmark & 74.85 & 76.32 & 86.77 & 0.8681 &  & 82.82 & 74.05 \\
w/o saliency map & \checkmark & 74.88 & 76.49 & 86.84 & 0.8690 &  & 82.57 & 73.82 \\
w/o sharpen & \checkmark & 75.54 & 77.01 & 87.15 & 0.8817 &  & 83.35 & 74.58 \\
w/o multiscale & \checkmark & 75.00 & 76.16 & 86.75 & 0.8745 &  & \textbf{84.64} & \textbf{76.05} \\
full & \checkmark & \textbf{77.22} & \textbf{78.01} & \textbf{87.97} & \textbf{0.8948} &  & 83.67 & 74.92 \\
\bottomrule
\end{tabular}
\end{center}
% \vspace{-0.3cm}
\end{table*}

\subsection{Comparison Study}
We compare our method with other methods in two training settings (supervised, semi-supervised). 
In the supervised setting, we select a subset from the whole training set described in \ref{data_datail} to make up the labelled training dataset (2,500 CAP slices and 2,500 NP slices from our private dataset, as well as 2,815 COVID-19 slices from the open dataset).
In the labelled training dataset, both CAP and NP slices have patient-level category labels and all zero segmentation masks since they are not COVID-19 cases, while all COVID-19 slices have patient-level category labels and pixel-level infection labels.
As for the semi-supervised setting, in addition to the labelled training dataset, we use the rest of the data in the full training set described in \ref{data_datail} without any labels to serve as the unlabelled training dataset for training.
Both supervised and semi-supervised settings share the same test datasets, 7 COVID-19 cases (1,336 slices) with pixel-level infection labels for segmentation performance test, as well as 187 cases (32,065 slices, NP: 60 cases, CAP: 60 cases, COVID-19: 60 private cases and 7 open cases) with patient-level category labels for classification performance test.
See more details about data split in Tabel~\ref{data_split}.
\par

For classification performance comparison, we consider four supervised methods: VGG-16, ResNet-18, JCS \cite{wu2021jcs}, Hu et al. \cite{hu2020weakly}, and two semi-supervised methods: Pseudo-Label \cite{lee2013pseudo}, MultiMix \cite{haque2021generalized}.
We use the classification performance of VGG-16 to serve as the supervised classification baseline.
For segmentation performance comparison, we consider three supervised methods: SegNet, U-Net, JCS \cite{wu2021jcs}, and two semi-supervised methods: PseudoSeg \cite{zou2020pseudoseg}, MultiMix \cite{haque2021generalized}.
We employ the segmentation performance of SegNet as the supervised segmentation baseline.
\par

All detailed quantitative classification and segmentation comparison performances are reported in Table~\ref{classify_performance} and Table~\ref{segmentation_performance}, respectively.
We also provide some qualitative visual comparisons of all methods in Fig.~\ref{fig:segs}.
In the supervised setting, our method achieves comparable classification performance with supervised counterparts (JCS, ResNet-18, and Hu et al.) and surpasses the supervised baseline (VGG-16) in all metrics.
Our method also gain the best segmentation performance compared with all supervised methods and the supervised baseline in both dice and mIoU. 
In the semi-supervised setting, our method still achieves superior classification (75.13\% accuracy, 76.35\% sensitivity, 86.89\% specificity, and 0.8748 AUC) and segmentation results (85.49\% dice, 76.97\% mIoU) compared with all the other methods and baselines.
Additionally, from Fig.~\ref{fig:segs}, we can get a more straight observation of the segmentation performance.
In contrast to other methods, our method shows superior segmentation results in both supervised and semi-supervised settings.
These quantitative and qualitative results demonstrate that our method is able to make good use of the unlabelled dataset and obtain superior performance with limited labelled data.

\begin{figure}[!b]
    \centering
    \includegraphics[width=.8\columnwidth]{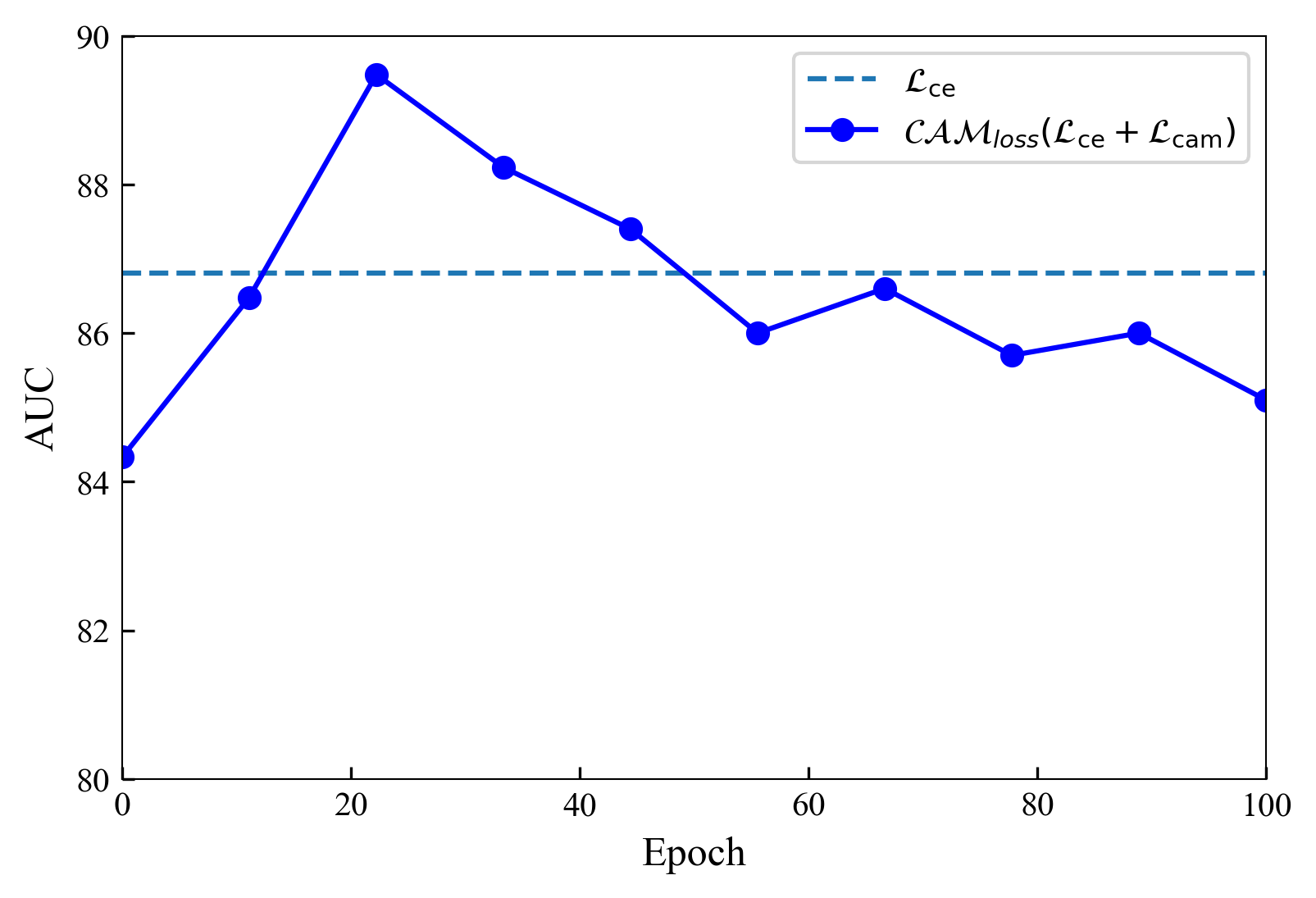}
    \caption{Ablation study of timing by using $\mathcal{CAM}_{loss}$ in the first 100 epochs. The dashed line indicates the performance of only using $\mathcal{L}_{\mathrm{ce}}$ for the classification task.}
    \label{fig:cam_loss_add_time}
\end{figure}

\subsection{Ablation Study}
To validate the effectiveness of every proposed component in our method, we perform an extensive ablation study.
Concretely, we remove the $\mathcal{CAM}_{loss}$ (only keep $\mathcal{L}_{\mathrm{ce}}$ for classification task), saliency map, sharpen combination module, and the multiscale learning strategy from our method, respectively (expressed as w/o in Table~\ref{ablation_results}).
Then we test their performance in both semi-supervised and weakly-supervised settings.
With regard to the weakly-supervised setting, we add patient-level category labels to the unlabelled training dataset for training and want to know whether patient-level category labels can provide performance gain or not.
\par

The ablation study results are reported in Table~\ref{ablation_results}.
As we can see, every component does contribute to the performance, especially in the semi-supervised setting.
The full version of our method achieves the best classification and segmentation results in all metrics.
The results under the weakly-supervised setting also illustrate a similar conclusion.
Whereas there is a little difference in the segmentation performance.
We can observe that, when adding the patient-level category label in the unlabelled training dataset, the full version of our method still obtains the best classification performance but achieve slightly lower segmentation performance compared with the w/o multiscale version.
We deem that this is caused by the coarse patient-level category labels.
In COVID-19 cases, not all slices are infected, thus there are no infections in this kind of slices.
These slices are actually as `healthy' as NP and CAP cases but assigned with COVID-19 category labels.
This brings noisy labels to our model and influences the performances.
While the number of this kind of slice is not large for the classification task, we can see the classification performance is still the best.
However they can be fatal for segmentation tasks since the number of labelled infections is limited and the quality of pseudo label relies on the saliency map and CAAM$_{l}$, which are controlled by the classification results.
The visual comparisons in Fig.~\ref{fig:segs} also show that the segmentation results under weakly-supervised settings are inferior to those under supervised or semi-supervised settings,
which confirms our deduction.
\par

To further support the results in Table~\ref{ablation_results},
we also plot the visualizations of the CT images and the corresponding saliency maps, CAAM$_{l}$, vanilla CAM, decoder prediction, pseudo label as well as the ground truth in Fig.~\ref{fig:maps_all}.
As illustrated, by fusing saliency map, CAAM$_{l}$ and decoder prediction through the sharpen combination module, we can obtain reliable pseudo labels. 
Moreover, $\mathcal{CAM}_{loss}$ does bring benefits in explainability, we can see a clear improvement from vanilla CAM to CAAM$_{l}$ and the activation areas match the infection areas in ground truth well.
Besides, we want to discuss more about the bad cases, as we can see in the bottom two rows, our model is not able to segment the infection areas. 
The main reasons here are twofold in these kinds of cases: firstly, the lung or the infection areas are too small for our model to produce accurate segmentation predictions and pseudo labels; secondly, the number of such kind of peripheral slices is quite limited during training procedure.
\par

Lastly, we look more closely at the timing of using $\mathcal{CAM}_{loss}$, i.e., finding the proper time to add $\mathcal{L}_{\mathrm{cam}}$ to $\mathcal{L}_{\mathrm{ce}}$.
As shown in Fig.~\ref{fig:cam_loss_add_time}, the most suitable timing is the $20^{th}$ epoch, where the model gains the best AUC.
At that moment, the vanilla CAM has already learnt obvious target category features, and it is a good time to use $\mathcal{CAM}_{loss}$ to enforce intra-class compactness and inter-class separability and refine the activation areas.

\section{Conclusion}
In this paper, we propose a semi-supervised multitask explainable identification and delineation method for COVID-19 infections, producing classification, segmentation and explainable visualizations at the same time.  
We demonstrate the effectiveness of our method with the combination of limited labelled data and unlabelled data or weakly-labelled data.
The key behind the performance gain is two-fold: Firstly, the $\mathcal{CAM}_{loss}$ helps to enforce intra-class compactness and inter-class separability, further bringing benefits in explainability. Secondly, we design a calibrated pseudo-labelling strategy and apply it under the consistency regularization framework, which only uses limited infection labels.
\par

Further exploration of patient-level category label issue, strong data augmentation method, the overall modest classification results caused by domain gap among centres worth more study as future directions.

\bibliographystyle{IEEEtran}
\bibliography{refs}

\end{document}